\documentclass[]{pasj02} 
\usepackage[switch,mathlines]{lineno} 
\usepackage{bm}
\usepackage{comment}
\usepackage{color}
\usepackage{url}
\usepackage{threeparttable}
\usepackage{rotating}

\jyear{2024}
\Received{}
\Accepted{}

\begin{document} 

\title{Magnetic field diagnostics of a solar active region filament}
\author{
  Daiki \textsc{Yamasaki},\altaffilmark{1}$^{,\dag}$\orcid{0000-0003-1072-3942}
  Yu Wei \textsc{Huang},\altaffilmark{2}
  Yuki \textsc{Hashimoto}, \altaffilmark{2}
  Satoru \textsc{UeNo},\altaffilmark{2}
  and
  Kiyoshi \textsc{Ichimoto}\altaffilmark{2}}
\altaffiltext{1}{Institute of Space and Astronautical Science, Japan Aerospace Exploration Agency, 3-1-1 Yoshinodai, Chuo-ku, Sagamihara, Kanagawa 252-5210, Japan}
\altaffiltext{2}{Astronomical Observatory, Kyoto University, Kitashirakawaoiwake-cho, Kyoto 606-8502, Japan}
\footnotetext[$\dag$]{Present address: yamasaki.daiki@jaxa.jp}

\KeyWords{Sun: filaments, prominences --- Sun: magnetic fields --- Sun: photosphere}

\maketitle

\begin{abstract}
  We performed spectropolarimetric observations of an active region filament in He I 10830 \AA~and Si I 10827 \AA~lines to investigate its magnetic field structure. We carried out full-Stokes inversions with the HAZEL code, which takes into account the Zeeman and Hanle effects. As a result, we yielded a mean field strength of 101 $\pm$ 33 G and a horizontal field nearly parallel to the filament axis, such that the distinction between the two classical normal- and reverse-polarity models becomes physically insignificant. In addition, we found Zeeman-like signatures in the linear polarization, characterized by double-peaked symmetric profiles, in some pixels of our observations. Since these profiles could not be well reproduced by modeling that included both the Zeeman and Hanle effects, we performed inversions assuming only the Zeeman effect. The inversion yielded a strong magnetic field of approximately 500 G. However, simultaneous observations of Si I 10827 \AA~indicate a photospheric magnetic field weaker than 100 G. Therefore, the scenario proposed by \citet{DiazBaso2016}, in which the strong field inferred from He I 10830 \AA~originates from contamination by the underlying photosphere, does not apply to our filament. The Zeeman-like profiles are preferentially found in optically thick regions ($\tau \sim$ 1.4--2.5), where the simplifying assumptions adopted in HAZEL are expected to become less reliable. Our results suggest that these profiles reveal limitations of the current inversion framework in optically thick regions and motivate future radiative-transfer modeling incorporating self-consistent radiation fields, differential illumination of the multiplet components, and possibly partial frequency redistribution.
\end{abstract}


\section{Introduction}
\begin{figure*}[htb]
  \begin{center}
    \includegraphics[width=0.8\linewidth]{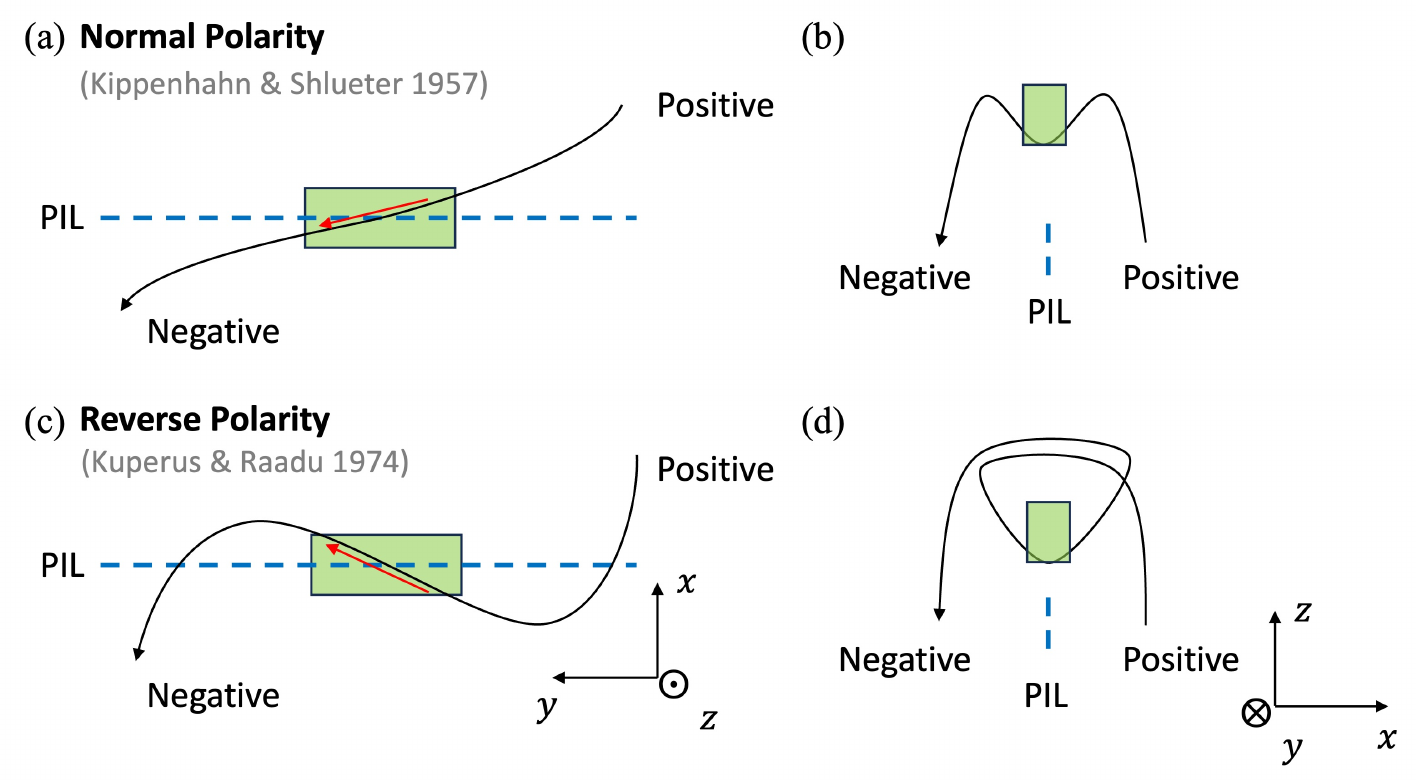}
  \end{center}
  \caption{Schematics of 2 classical models of filament magnetic field configuration. (a) Top view of normal polarity model. (b) Side view along the filament axis of normal polarity model. (c) Top view of reverse polarity model. (d) Side view along the filament axis of reverse polarity model. Black arrow, green box, and blue dashed line correspond to magnetic field, plasma material of the filament, and polarity inversion line, respectively. \textbf{Alt text: Four schematic diagrams comparing the normal-polarity and reverse-polarity filament models. Top and side views in the left and right illustrate the magnetic field orientation relative to the polarity inversion line and filament axis. In the normal-polarity model in upper panels, the magnetic field direction follows the photospheric polarity pattern, while in the reverse-polarity model in lower panels the field direction is reversed.}}\label{fig1}
\end{figure*}

Solar dark filaments (or prominences when seen on the solar limb) are cool ($\sim10^{4}$ $\mathrm{K}$) and dense ($>10^{9}$ $\mathrm{cm^{-3}}$) plasma structures suspended in the hot ($>10^{6}$ $\mathrm{K}$) corona.
They are observed as dark absorption features in the H$\alpha$ line and are thought to be supported by dipped magnetic field lines above the photospheric polarity inversion lines (PILs; \cite{Babcock1955}).
Depending on their magnetic and thermodynamic environments, they are generally classified into quiescent-region (QS) filaments and active-region (AR) filaments.
While QS filaments are long-lived and located in weak-field regions, AR filaments form in strong-field, highly sheared magnetic environments and often become unstable, leading to filament eruptions, solar flares, and coronal mass ejections (CMEs; \cite{Priest2002, Shibata2011, Parenti2014}).
\\
~
Understanding the magnetic field structure that supports filaments is essential to clarify the mechanisms governing their stability and eruption.
Two classical models have been proposed for the magnetic configuration of filaments: the normal-polarity model \citep{Kippenhahn1957}, in which the filament magnetic shear is aligned with the photospheric magnetic field, and the reverse-polarity model \citep{Kuperus1974}, in which the shear direction is opposite (see Figure \ref{fig1}).
Previous observations have shown that QS filaments typically exhibit reverse polarity ($e.g.,$ \cite{Bommier1998}; \cite{Martinez2015}; \cite{Wang2020}; \cite{Yamasaki2023}), whereas both normal and reverse configurations have been reported for AR filaments \citep{Xu2012,Sasso2014,Yokoyama2019}.
\newpage
\noindent
Diagnosing the vector magnetic field in filaments requires spectropolarimetric measurements of magnetically sensitive chromospheric lines.
Among them, the He I 10830 \AA~triplet is particularly powerful because it is sensitive to both the Zeeman \citep{Zeeman1897} and Hanle effects \citep{Hanle1924, TrujilloBueno2002}, allowing quantitative determinations of field strength and orientation over a wide dynamic range \citep{Yamasaki2023,Hashimoto2026}.
However, the interpretation of He I 10830 \AA~polarization is not straightforward.
\citet{VicenteArevalo2023} demonstrated that, in filaments with moderate to large optical thickness, radiative transfer and non-uniform illumination in wavelength over the three components of the He I 10830 \AA~multiplet have non-negligible effects in shaping the He I 10830 \AA~polarization, and that neglecting these effects can lead to significant errors in the inferred magnetic field vector.
\\
~
Previous work by \citet{Yamasaki2023} analyzed eight QS filaments using He I 10830 \AA~spectropolarimetric data obtained with the Domeless Solar Telescope (DST: \cite{Nakai1985}) at Hida Observatory.
They found field strengths of 8–35 $\mathrm{G}$ and concluded that most of the filaments had reverse-polarity configurations.
\citet{Hashimoto2026} observed one of the target filaments of \citet{Yamasaki2023} as an off-limb prominence and reported the consistent results on magnetic field strength and the direction.
In contrast, AR filaments are expected to host much stronger magnetic fields, and \citet{Kuckein2009} reported Zeeman-like Stokes profiles showing symmetric linear polarization without clear Hanle signatures, from which magnetic field strengths of several 100 G were inferred.
Other observational studies also reported magnetic field strengths of several 100 G in AR filaments \citep{Sasso2011,Xu2012,Sasso2014}.
\citet{DiazBaso2016} proposed that such profiles of He I 10830 \AA~inferring a strong magnetic field can be reproduced by a two-layer model consisting of a strong-field lower component and a weak-field upper component along the line of sight, and the Zeeman-like profiles are formed in the lower layer.
\\
~
In this study, we extend the spectropolarimetric analysis of \citet{Yamasaki2023} to an active-region filament located above NOAA 13092, observed on 2022 September 5 with the DST.
By applying the HAZEL inversion code \citep{AsensioRamos2008} to the full Stokes profiles of the He I 10830 \AA~triplet, we derived the magnetic field strength and configuration of the filament.
We also performed inversions of the Si I 10827 \AA~line to compare the photospheric and chromospheric magnetic fields and examined the 180 deg azimuth ambiguity using two different resolution methods.
\\
~
The rest of this paper is structured as follows; the observation is introduced in Section \ref{sec:obs}, the analysis is described in Section \ref{sec:ana}, results are presented in Section \ref{sec:res}, and discussions on our findings are given in Section \ref{sec:dis}.

\section{Observation} \label{sec:obs}
\begin{figure*}[htb]
  \begin{center}
    \includegraphics[width=0.7\linewidth]{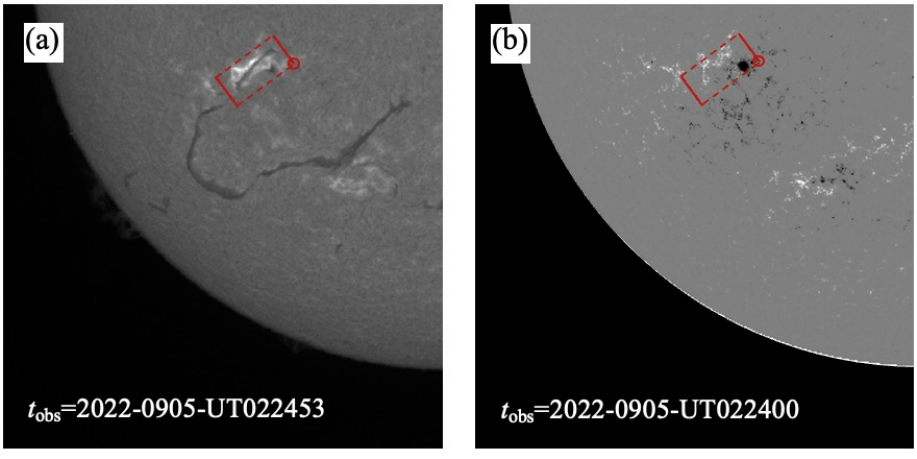}
  \end{center}
  \caption{Location of the observational targets taken on 2022 Sep. 5. (a) H$\alpha$ line center image taken by the SMART-SDDI. (b) Photospheric magnetogram taken with SDO-HMI. Greyscale shows the radial component of the magnetic field in the range of $\pm750$ G. Red rectangles show the field-of-view of the target region. Solid and dashed lines indicate the slit and scan directions, respectively. Circle symbol represents the origin of the field of view. \textbf{Alt text: Two solar images showing the location of the target active-region filament observed on 2022 September 5. The left panel shows an H$\alpha$ image with the filament appearing as a dark elongated structure. The right panel shows the photospheric magnetic field map with positive and negative polarities surrounding the filament. Red rectangles indicate the spectropolarimetric observation field of view.}}\label{fig2}
\end{figure*}

Our target active region (AR) filament was located in the AR NOAA 13092.
On September 5, we performed the spectropolarimetric observations using the spectro-polarimeter \citep{Ichimoto2022, Yamasaki2022a} on the Domeless Solar Telescope (DST) at Hida Observatory.
He I 10830 \AA~and Si I 10827 \AA~lines were simultaneously taken with an exposure time of $15$ $\mathrm{msec}$ for each frame.
We obtained $200$ frames in $3$ $\mathrm{sec}$ for each slit position with $1.0$ $\mathrm{Hz}$ rotation of the rotating waveplate. 
The slit width and length were $0.1$ and $20$ $\mathrm{mm}$, corresponding to $0''.64$ and $128''$ on the sky, respectively.
Spatial sampling along the slit is $0''.43$ $\mathrm{pixel^{-1}}$, and the spectral sampling is 29 m\AA~$\mathrm{pixel^{-1}}$ while scan step was $1''.38$.
The total number of slit positions was 200, thus the the field of view (FOV) in scan direction is $276''$, which took about 10 minutes for whole target scan.
\\
~ The observation target is indicated on a H$\alpha$ solar image taken by the Solar Dynamics Doppler Imager (SDDI: \cite{Ichimoto2017}) on the Solar Magnetic Activity Research Telescope (SMART: \cite{UeNo2004}) in Figure \ref{fig2} (a).
In addition, we also show an image of the photospheric magnetic field taken by the Helioseismic and Magnetic Imager (HMI: \cite{Scherrer2012}) onboard the $Solar~ Dynamics~ Observatory$ ($SDO$: \cite{Pesnell2012}) in Figures \ref{fig2} (b).
We note that a few hours before our spectro-polarimetric observation, a GOES C-class flare occurred in the target AR.
\\
~ Geometrical height of the target filament from the solar surface is an important parameter for a correct evaluation of the magnetic field by using the Hanle effect.
In our study, we estimated the height of the target filament using the SMART-SDDI images on August 31, $i.e.,$ 5 days prior to the DST spectro-polarimetric observation when the prominence was observed on the limb.

\section{Analysis}\label{sec:ana}
\subsection{Data reduction and calibration}
The obtained spectral data were reduced with dark frame subtraction and flat-fielding.
The orthogonally polarized spectra recorded simultaneously by the detector were aligned after correcting the distortion of spectral images by referring to photospheric absorption lines and 5 hair lines imaged on the spectra taken in the quiet sun after the observation sequence.
Then we performed polarization demodulation for each set of 200 images, and combined them to obtain Stokes $I$, $Q$, $U$, and $V$ spectrum. 
Regarding the calibration of the instrumental polarization of the DST, we referred to \citet{Anan2018}.
For an initial guess of the M\"{u}ller matrix of the DST, we used the data obtained by \citet{Anan2012}.
Further adjustment of the matrix is done by using the sunspot observation data (cf. \cite{Ichimoto2022}).
After polarization calibration, we removed artificial fringe pattern in Stokes $Q$, $U$, and $V$ spectra by subtracting periodic component in dispersion direction.
For more details of the calibration procedure, see also \citet{Ichimoto2022} and \citet{Yamasaki2023}.
\\
~
In Figure \ref{fig3}, we show the distribution of the Stokes signals in He I 10830 \AA~over the dark filament.
All the Stokes signals are normalized by the continuum intensity.
Panel (a) displays the intensity at the line center of He I 10830 \AA, where we can identify the target filament as a dark structure.
Green contour shows the border of the filament determined from the Stokes $I$ map and the interior of which is used to extract the target region.
Regarding the linear polarization signals of Stokes $Q$ and $U$ in panels (b) and (c), we display the integrated values in range of $\pm0.36$ \AA~around the center of the red component of He I 10830 \AA.
Regarding the circular polarization signal of Stokes $V$ in panel (d), we display the value of the subtraction of the integrated red and blue wings of the red component in range of $+0.36\pm0.14$ \AA~and $-0.36\pm0.14$ \AA, respectively.
We find positive signals of Stokes $V$ over the filament body (see panel (d) in Figure \ref{fig3}). 
Note that the panel (a) for the intensity shows the observed full FOV, while in panels (b)-(f) show a smaller FOV including the filament.
Figure \ref{fig3} (e) shows the spatial distribution of the Doppler velocity derived from the He I 10830 \AA.
Along the filament axis, blueshifted and redshifted regions are observed on opposite sides of the filament.
The magnitude of these Doppler shifts is about $\pm 10~\mathrm{km/s}$.

\subsection{Stokes Inversion}\label{subsec:inv}
To obtain the magnetic field in the filament, we performed the Stokes inversion for full Stokes profiles of He I 10830 \AA~by the HAZEL code, which takes into account the atomic polarization as well as the Zeeman effect.
HAZEL is based on the assumption of complete frequency redistribution (CRD), the slab approximation, and the flat-spectrum approximation \citep{AsensioRamos2008,VicenteArevalo2023}. 
HAZEL obtains 8 physical parameters from the fitting of observed Stokes profiles, $i.e.,$ the magnetic field strength ($|\bm{B}|$), the inclination ($\theta$) and azimuth ($\phi$) of the magnetic field vector with respect to the local vertical, the optical thickness ($\tau$), the Doppler velocity ($v_{\mathrm{doppler}}$), the turbulent velocity ($v_{\mathrm{turb}}$), the line damping ($a$), and the filling factor ($ff$).
\\
~
For the whole area of the dark filament, we performed Stokes inversion using the HAZEL code, which fully takes into account the Zeeman effect, the population imbalance, and quantum coherences of the He I levels induced by the anisotropic radiation field allowing scattering polarization and the Hanle effect to contribute to the emergent Stokes profiles.
Not to miss possible solutions due to the Van-Vleck ambiguity (\cite{LandiDegl'Innocenti2004}, \cite{AsensioRamos2008}), we performed the fitting for 3 different ranges of the inclination angle, $\theta$ in the local frame separately.
They are referred to ``case A'' for $0^{\circ}.0<\theta<54^{\circ}.74$, ``case B'' for $54^{\circ}.74<\theta<125^{\circ}.26$, and ``case C'' for $125^{\circ}.26<\theta<180^{\circ}.0$ in this paper, where $54^{\circ}.74$ is the so called Van-Vleck angle.
Regarding the other physical parameters, we set the range of their values as $0.0<|\bm{B}|<1000.0$ $\mathrm{Gauss}$, $0^{\circ}.0<\phi<360^{\circ}.0$, $0.1<\tau<5.0$, $-20.0<v_{\mathrm{doppler}}<20.0$ $\mathrm{km/s}$, $3.0<v_{\mathrm{turb}}<15.0$ $\mathrm{km/s}$, $0.0<a<1.0$, and fixed $ff=1.0$.
For some particular pixels, in which we observed Zeeman-like profiles in Stokes profiles, we applied the HAZEL inversion with a reduced atomic polarization following \citet{DiazBaso2016}.
In this case, the contribution of atomic level polarization was artificially suppressed, resulting in polarization signals dominated by the Zeeman effect.
\\
~ 
To deduce the photospheric vector magnetic field, we applied the Milne Eddington inversion to the full Stokes profiles of Si I 10827 \AA~line.
The inversion code includes 8 physical parameters: the magnetic field strength, the inclination and azimuth of the magnetic field vector in observer's frame, the line of sight velocity, the turbulent velocity, the damping constant of the line, the continuum intensity, and the ratio of the opacities in line center and continuum.

\subsection{Resolving 180 deg ambiguity}\label{subsec:180}
Regarding the He I 10830 \AA, taking into account the projection effect and the sign of the longitudinal magnetic field in observer's frame ($B_{\mathrm{l}}$), we solved the $180$ degree ambiguity problem ($cf.$ \cite{Yamasaki2023}).
As an illustrative example, let us consider a filament whose axis is approximately aligned with the line connecting the disk center and the limb on the solar disk.
If the filament exhibits a positive Stokes $V$ signal over most of its length, the line-of-sight component of the filament magnetic field is directed toward the observer.
In this geometry, under the assumption that the filament thread patterns, which are obviously elongated horizontally with respect to the solar surface, represent the direction of the filament magnetic field.
It implies that the field is likely oriented from the limb toward the disk center, rather than from the disk center toward the limb.
Therefore, the sign of Stokes $V$ provides an additional criterion to select the correct azimuthal direction of the transverse field and thus helping to resolve the 180 deg ambiguity.
\\
~
To resolve the 180 deg ambiguity of photospheric vector magnetic field obtained from Si I 10827 \AA~observation, we performed 2 different methods.
One is the potential field method, which selects the horizontal magnetic field orientation that has an acute angle to the potential field derived from the longitudinal magnetic field.
The other is the minimum energy method \citep{Metcalf1994}, which resolves the 180 deg ambiguity by minimizing a cost function that penalizes both non-zero magnetic divergence and excessive electric currents, thereby selecting a field configuration that approximately satisfies divergence free and has minimum free magnetic energy.

\clearpage
\begin{sidewaysfigure*}[htb]
  \begin{center}
    \includegraphics[bb= 0 0 845 325, width=200mm]{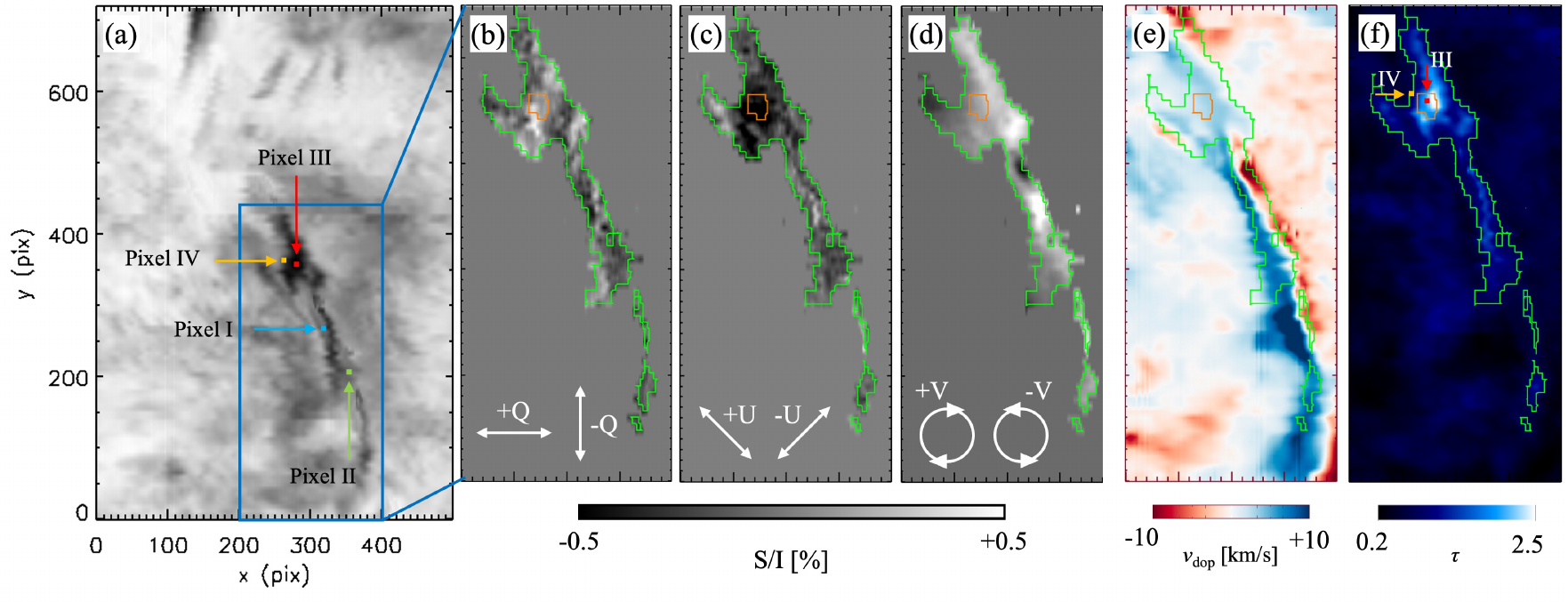}
  \end{center}
  \caption{Distribution of Stokes signal in He I 10830 \AA~in the target filament. (a) Stokes I image at the line center of He I 10830 \AA~of the full field of view. The cyan, green, red, and yellow symbols in the panel show pixels I, II, III, and IV, respectively. The Stokes profiles in these pixels I-IV will be described in detail in subsequent sections. The blue box corresponds to the region of interest (ROI), which includes the target filament. (b, c) Stokes Q and U map of ROI. The displayed Stokes signals are integrated values in the range $\pm0.36$ \AA~around the line center of He I 10830 \AA~and normalized by the continuum intensity. (d) Stokes V map of ROI. The displayed Stokes signal is made from the subtraction of the red ($+0.36\pm0.14$ \AA) and blue ($-0.36\pm0.14$ \AA) wings normalized by the continuum intensity. (e) Doppler velocity obtained from He I 10830 \AA. (f) Optical thickness of He I 10830 \AA. The green contour shows the mask for the target dark filament region. Orange contour indicates the location of the Zeeman-like profiles in Stokes $U$. \textbf{Alt text: Six panels including 4 maps of the He I 10830 \AA~Stokes parameters, Doppler velocity, and optical thickness over the target filament. The filament appears as a dark elongated structure with positive Stokes V signals along most of its body. Blueshifted and redshifted regions are distributed on opposite sides of the filament, suggesting a counter-streaming flow. Regions showing Zeeman-like polarization profiles are highlighted by orange contours.}}\label{fig3}
\end{sidewaysfigure*}
\clearpage

\section{Results}\label{sec:res}
\subsection{Overview of the magnetic field structure of the target filament}\label{subsec:invres}
\begin{figure*}[htb]
  \begin{center}
    \includegraphics[width=0.7\linewidth]{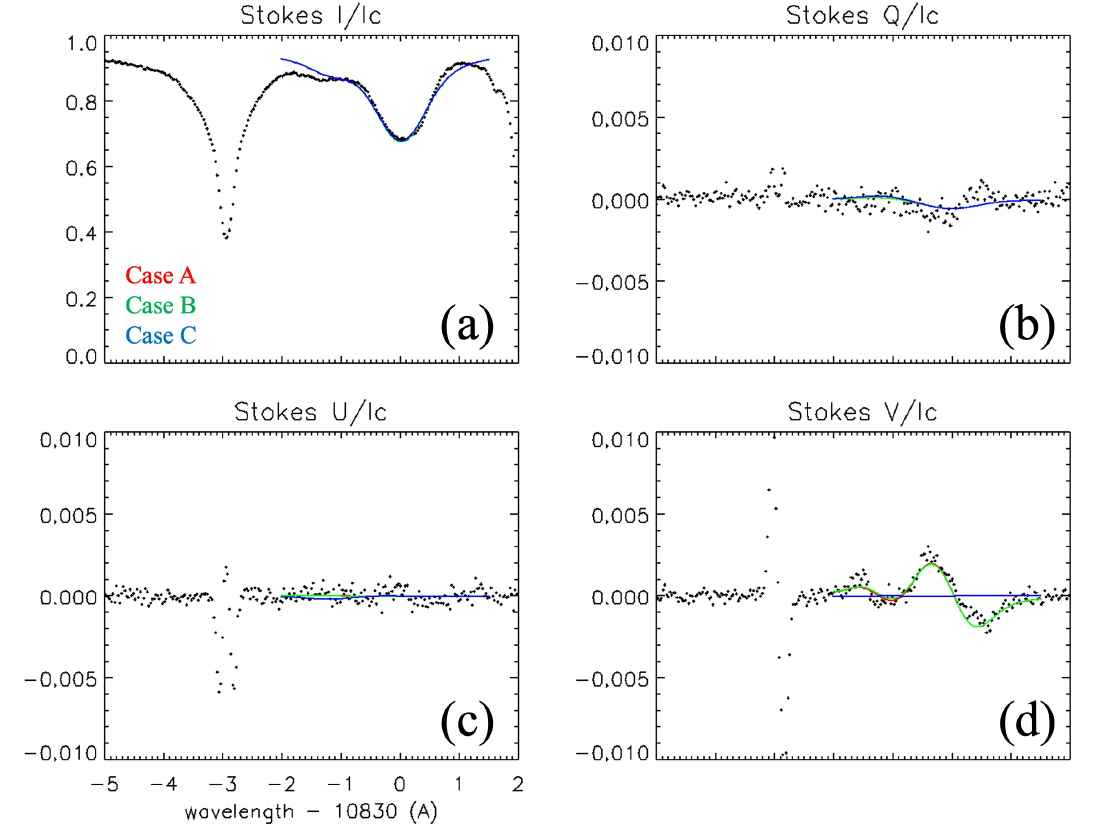}
  \end{center}
  \caption{Stokes profiles of I, Q, U, and V at pixel I (see the cyan symbol in Figure \ref{fig3} (a)). The black symbols show the observational data. The red, green, and blue solid lines show the full atomic polarization fitting results for ``case A,'' ``case B,'' and ``case C,'' respectively. Note that the continuum level used in normalization of intensity is determined by that of the disk center. \textbf{Alt text: Four panels of observed and modeled Stokes I, Q, U, and V profiles at pixel I within the filament. Three inversion solutions corresponding to different magnetic inclination ranges are overplotted. Cases A and B reproduce the observed Stokes V profile similarly well, while case C fails to fit the circular polarization signal.}}\label{fig4}
\end{figure*}

\begin{figure*}[htb]
  \begin{center}
    \includegraphics[width=0.7\linewidth]{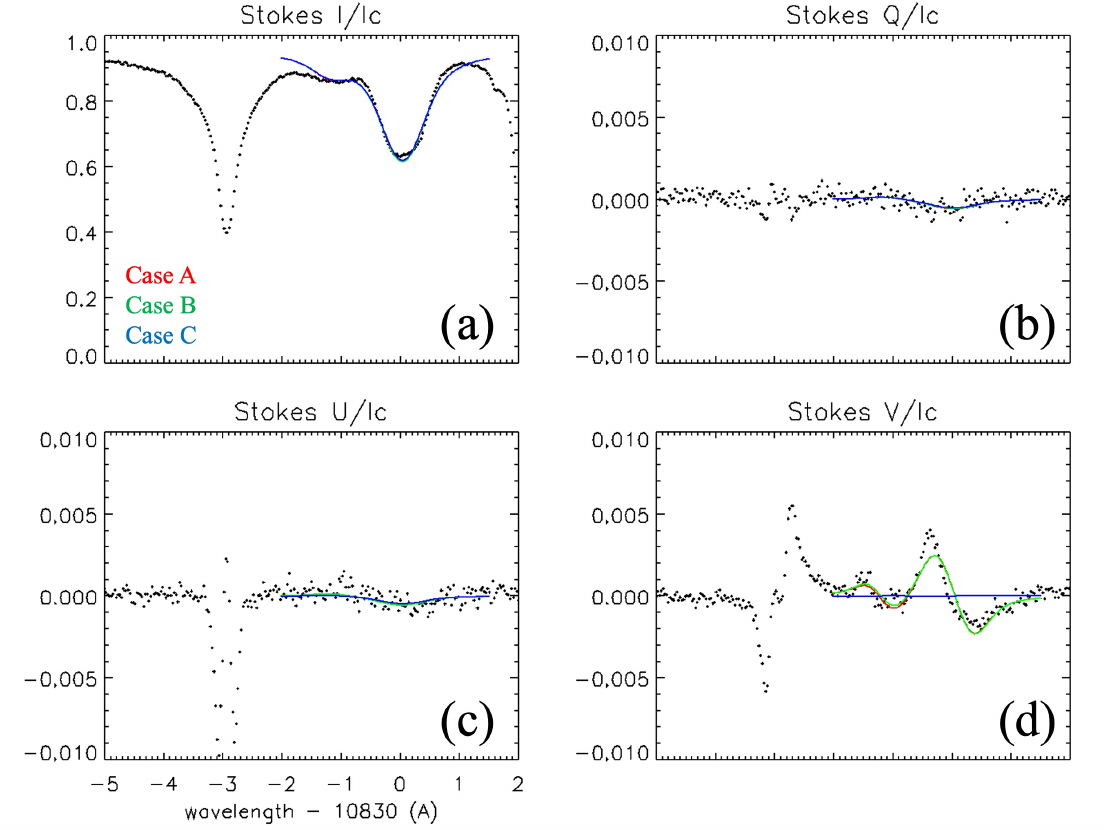}
  \end{center}
  \caption{Stokes profiles of I, Q, U, and V at pixel II (see the green symbol in Figure \ref{fig3} (a)). The black symbols show the observational data. The red, green, and blue solid lines show the full atomic polarization fitting results for ``case A,'' ``case B,'' and ``case C,'' respectively. Note that the continuum level used in normalization of intensity is determined by that of the disk center. \textbf{Alt text: Four panels of observed and modeled Stokes profiles at pixel II in the filament. The profiles are generally reproduced by cases A and B of the full atomic polarization inversion, while case C does not reproduce the observed Stokes V signal.}}\label{fig5}
\end{figure*}

\begin{figure*}[H]
  \begin{center}
    \includegraphics[bb= 0 0 442 543, width=140mm]{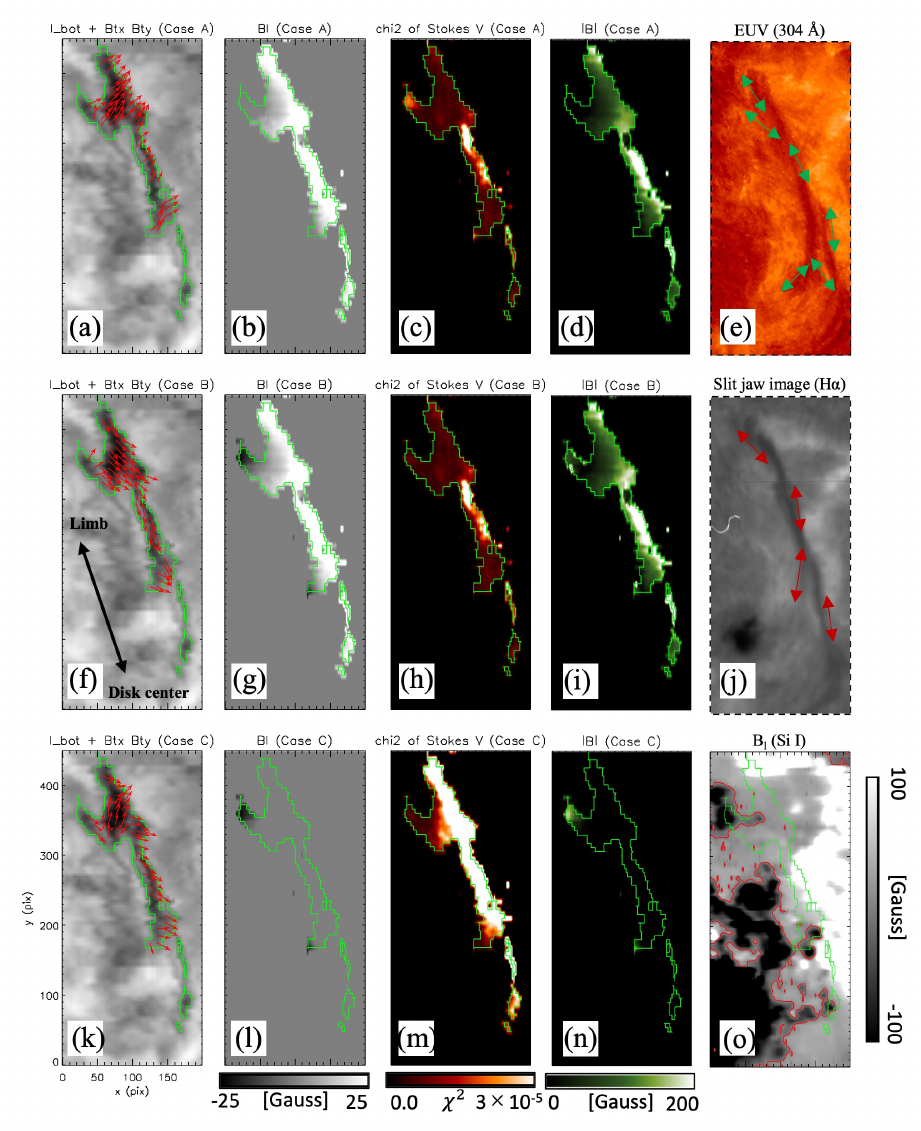}
  \end{center}
  \caption{Vector magnetic field of the target AR filament obtained from the HAZEL inversion. The green contour shows the border of the target dark filament region. (a) Transverse magnetic field vector in the observer's frame with red arrows overplotted on the Stokes I image at the line center of He I 10830 \AA~for ``case A.'' (b) Longitudinal magnetic field in the observer's frame for ``case A.'' (c) $\chi^{2}$ distribution of Stokes V for ``case A.'' (d) Magnetic field strength for ``case A.'' (e) EUV 304 \AA~image taken with AIA. The green arrows show the direction of the threads of the filament. (f) As panel (a) for ``case B.'' (g) As panel (b) for ``case B.'' (h) As panel (c) for ``case B.'' (i) As panel (d) for ``case B.'' (j) Slit-jaw image at the H$\alpha$ line center, 6563 \AA. The red arrows show the direction of the threads of the filament. (k) As panel (a) for ``case C.'' (l) As panel (b) for ``case C.'' (m) As panel (c) for ``case C.'' (n) As panel (d) for ``case C.'' (o) Line-of-sight component of the photospheric magnetic field obtained from Si I 10827 \AA~line observation. The white- and black-colored regions correspond to positive and negative polarities, respectively. Red contour shows the PIL. \textbf{Alt text: Fifteen panels are displayed. Twelve out of fifteen correspond to maps of the magnetic field structure derived from the HAZEL inversions for three inclination-angle cases; four panels for each case including the transverse and longitudinal magnetic field maps, fitting quality maps, magnetic field strength distributions. The rest three panels correspond to the photospheric magnetic field derived from Si I 10827 \AA~observation, EUV 304 \AA~and H$\alpha$ images for filament thread orientations comparison. The adopted case B solution shows horizontal magnetic fields nearly parallel to the filament axis.}
}\label{fig7}
\end{figure*}

In Figures \ref{fig4} and \ref{fig5}, we show the observed Stokes profiles and the result of the HAZEL inversion for pixels I and II, which represent the main body of the filament, respectively (see the symbols marked by cyan and green in Figure \ref{fig3} (a)).
Pixels I has the same signs of Stokes $V$ in Si I and He I, while pixel II has the opposite signs.
As shown in Figure \ref{fig3} (e), the target filament exhibits a line-of-sight velocity pattern suggesting counter-streaming flows.
The influence of such velocity fields, particularly in regions where the red- and blue-shifted components overlap, results in complex Stokes profiles.
We therefore selected representative pixels in the blue-shifted side for which the profiles are relatively simple and thus more amenable to interpretation.
In Figures \ref{fig4} and \ref{fig5}, the red, green and blue solid curves overlaid on He I 10830 \AA~profiles in each panel are the fitting results from the HAZEL inversion with the full atomic polarization for the ``case A'', ``case B'', and ``case C'', respectively.
As shown in panel (a) of Figures \ref{fig4} and \ref{fig5}, all the cases successfully fit the observed intensity profile.
Regarding the fitting of Stokes $V$ profiles shown in panel (d) of Figure \ref{fig4} and \ref{fig5}, we find similarly good fits in ``case A'' and ``case B'', while the fitting in ``case C'' fails.
The physical parameters at pixels I and II obtained from the inversion for three cases are summarized in Table \ref{tab1} and \ref{tab2}, respectively.
These results indicate that both ``case A'' and ``case B'' remain plausible solutions.
Determining the most appropriate solution requires consideration beyond a single-pixel analysis, and must instead be evaluated in the context of the global magnetic orientation of the filament and its consistency with complementary imaging observations (cf. \cite{Yamasaki2023}).
\\
~ In Figure \ref{fig7}, we show the results of full atomic polarization inversion over the target filament.
In the first, second, third, and fourth column, we show the transverse magnetic field vector in observer's frame, the longitudinal magnetic field in observer's frame, the $\chi^2$ map for the Stokes $V$, and the magnetic field strength ($|B|$) for the cases A, B, and C from the top to the bottom, respectively.
To resolve the Van-Vleck ambiguity, we compared the threads pattern in 304 \AA~image taken by AIA in panel (e) and H$\alpha$ image in panel (j) and the horizontal component of the vector magnetic field of ``case A'' and ``case B'' where the orientations of threads pattern are depicted by arrows in panels (e) and (j) for clarity.
We found that the ``case B'' result shows a similar direction to those images.
Then, we finally adopted the inversion result from ``case B'', in which we performed the inversion with the inclination angle restriction of $54^{\circ}.74<\theta<125^{\circ}.26$, as our solution.
Regarding the magnetic field strength, we obtained 101 $\pm$ 33 $\mathrm{G}$ (mean $\pm$ 1$\sigma$) for the target filament area from the ``case B'' results.
\\ 
~ 
As shown in panel (g) of Figure \ref{fig7}, $B_{\mathrm{l}}$ is positive over the most part of the filament body.
Therefore, the direction of the magnetic field of the filament is towards us, $i.e.$, the axial magnetic field is expected to be from solar limb to disk center.
In this way, we determined the direction of the transverse magnetic field in observer's frame ($\bm{B_{\mathrm{t}}}$) as shown in red arrows in panel (f) of Figure \ref{fig7}.

\begin{table}[htb]
   \begin{threeparttable}
    \caption{Inversion results for three cases at pixel I with full atomic polarization cases A, B, and C}
    \label{tab1}
    \begin{tabular}{lrrrrrrrr}
      \hline
      & $|\bm{B}|$ & $\theta$ & $\phi$ & $\tau$  & $v_{\mathrm{Dop}}$ & $v_{\mathrm{turb}}$ & $a$  & $ff$ \tnote{1}  \\
      & [G]     & [$^\circ$]    & [$^\circ$]  &        & [km/s]           & [km/s]            &        &        \\
      \hline
      A & $109$ & $43$  & $41$  & $0.6$ & $-8$ & $9$ & $1$ & $1$  \\
      B & $109$ & $66$  & $0$   & $0.6$ & $-8$ & $9$ & $1$ & $1$  \\
      C & $2$   & $142$ & $123$ & $0.6$ & $-8$ & $9$ & $1$ & $1$  \\
      \hline
    \end{tabular}
    \begin{tablenotes}
    \item[1] The filling factor was fixed at unity for all cases.
    \end{tablenotes}
  \end{threeparttable}
\end{table}

\begin{table}[htb]
\begin{threeparttable}
  \caption{Inversion results for three cases at pixel II with full atomic polarization cases A, B, and C}
    \label{tab2}
    \begin{tabular}{lrrrrrrrr}
      \hline
      & $|\bm{B}|$ & $\theta$ & $\phi$ & $\tau$  & $v_{\mathrm{Dop}}$ & $v_{\mathrm{turb}}$ & $a$    & $ff$ \tnote{1}  \\
      & [G]     & [$^\circ$]    & [$^\circ$]  &        & [km/s]           & [km/s]            &        &        \\
      \hline
      A & $87$ & $33$  & $38$ & $0.8$ & $-8$ & $7$ & $1$ & $1$  \\
      B & $93$ & $66$  & $13$ & $0.8$ & $-8$ & $7$ & $1$ & $1$  \\
      C & $1$  & $125$ & $73$ & $0.8$ & $-8$ & $7$ & $1$ & $1$  \\
      \hline
    \end{tabular}
    \begin{tablenotes}
      \item[1] The filling factor was fixed at unity for all cases.
    \end{tablenotes}
\end{threeparttable}
\end{table}

\subsection{Photospheric magnetic field}
\begin{figure}[htb]
  \begin{center}
    \includegraphics[width=\linewidth]{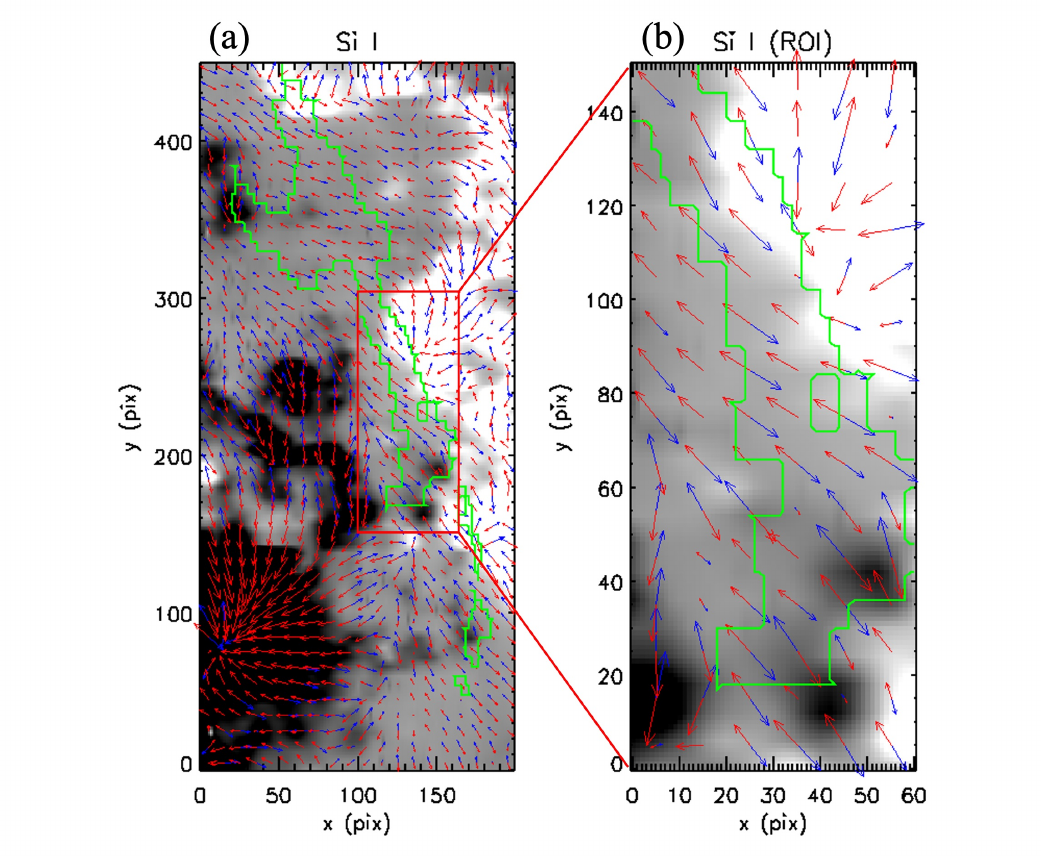}
  \end{center}
  \caption{Photospheric vector magnetic field obtained from Si I 10827 \AA~observation. Background greyscale shows line-of-sight component of the magnetic field. Red and blue arrows correspond to the horizontal component of the magnetic field from potential field method and minimum energy method, respectively. (a) Full field of view. (b) Region of interest, center part of the target filament. \textbf{Alt text: Two panels are displayed. One shows photospheric vector magnetic field maps for the whole field of view derived from Si I 10827 \AA~observations using two different 180-degree ambiguity resolution methods. Red and blue arrows indicate the transverse magnetic field directions obtained from the potential-field and minimum-energy methods, respectively. The other panel shows the enlarged view around the PIL. The methods agree in sunspot regions but differ near the polarity inversion line beneath the filament.}}\label{fig6}
\end{figure}

As we introduced in Section \ref{subsec:180}, we applied two different 180 deg ambiguity solvers on the photospheric vector magnetic fields: the potential field method and the minimum energy method.
In Figure \ref{fig6}, we show the photospheric vector magnetic field obtained from Si I 10827 \AA~observation with 2 different 180 deg ambiguity solvers.
Red and blue arrows show the transverse component of the photospheric magnetic field determined by the potential field method and the minimum energy method, respectively.
Within the sunspot umbra and penumbra, the two methods generally yielded consistent directions of the transverse magnetic field.
However, discrepancies were found in the regions near the magnetic polarity inversion line (PIL), $i.e.,$ beneath the filament, where the inferred azimuthal directions of the horizontal field are opposite.
This indicates that ambiguity resolution based solely on photospheric magnetic information is challenging in such regions.
On the other hand, when we consider the orientation of the horizontal magnetic field within the filament, we find that it is consistent with the result obtained using the Minimum Energy method.
Since highly sheared axial component of magnetic field has a same direction in filament and underlying photosphere irrespective of normal and reverse polarity configuration (Figure \ref{fig1}), this suggests that the minimum energy method selected the correct directions of magnetic field.
In this way, the transverse field derived in a filament can provide an independent and useful constraint for resolving the 180 deg ambiguity in the photospheric magnetic field.

\subsection{Zeeman-like profiles in He I 10830 \AA}\label{subsec:zeeman}
\begin{figure}[htb]
  \begin{center}
    \includegraphics[width=\linewidth]{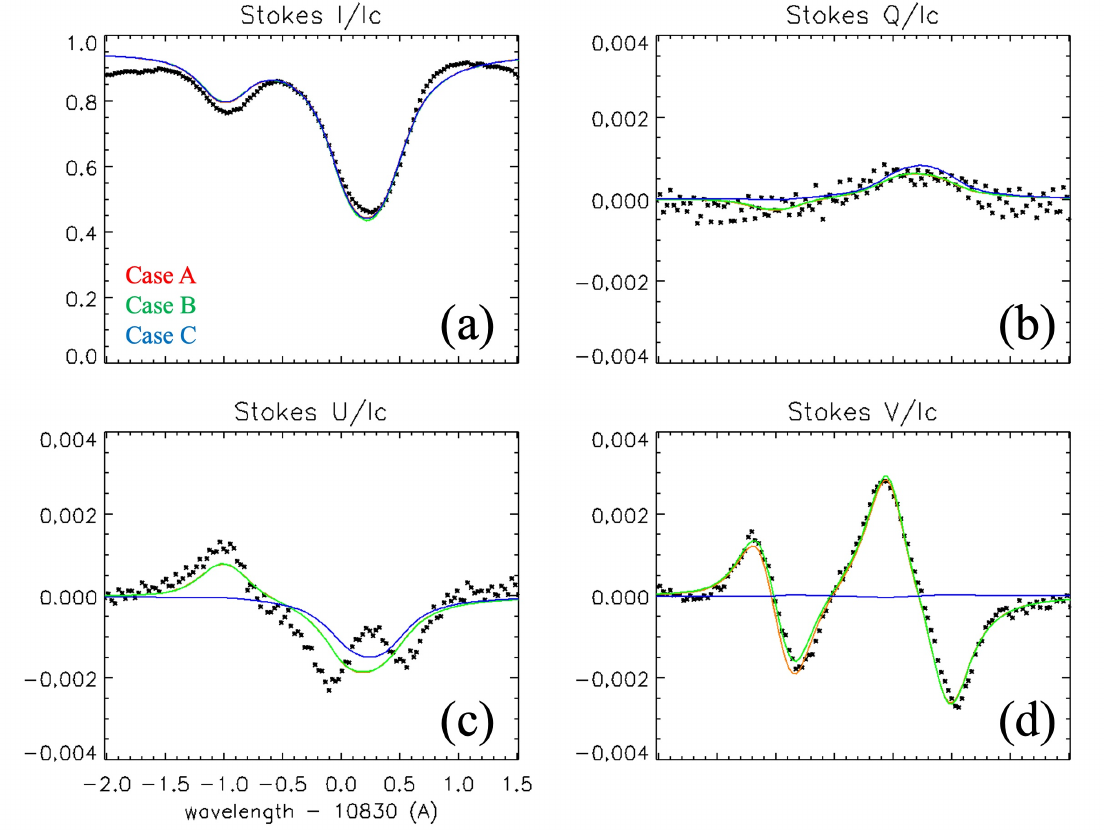}
  \end{center}
  \caption{Stokes profiles of I, Q, U, and V at pixel III (see the red symbol in figure \ref{fig3}a). The black symbols show the observational data. The red, green, and blue solid lines show the full atomic polarization fitting results for ``case A,'' ``case B,'' and ``case C,'' respectively. Note that the continuum level used in normalization of intensity is determined by that of the disk center. \textbf{Alt text: Four panels of observed and modeled Stokes profiles at pixel III, where a Zeeman-like polarization profile is detected. The Stokes U profile exhibits a clear double-peaked symmetric structure in the red component of He I 10830 Å. Full atomic polarization inversions cannot fully reproduce the observed profile shape.}}\label{fig8}
  \end{figure}

\begin{figure}[htb]
  \begin{center}
    \includegraphics[width=\linewidth]{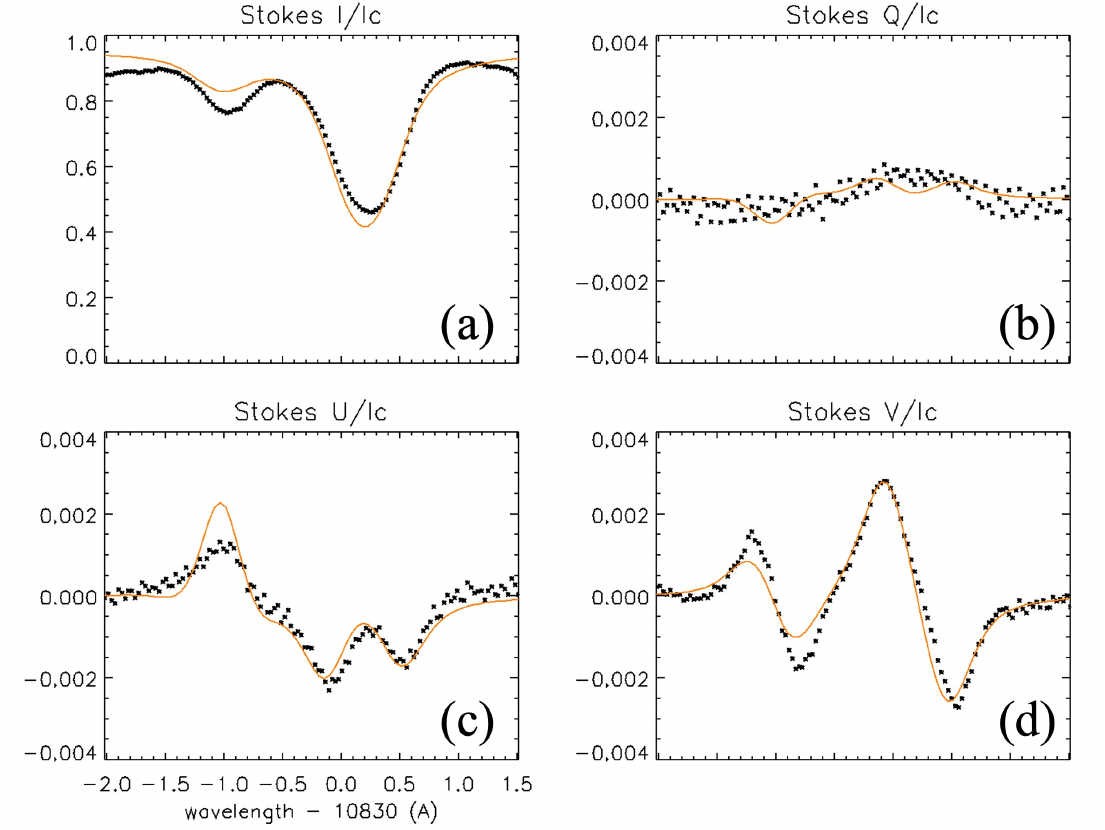}
  \end{center}
  \caption{Same as Figure \ref{fig8}, but for the reduced atomic polarization fitting result. The wavelength range is limited around He I 10830 \AA. \textbf{Alt text: Same as Figure \ref{fig8}. Observed Stokes profiles at pixel III fitted using the reduced atomic polarization inversion. The Zeeman-like double-peaked Stokes U profile is reproduced more successfully than in the full atomic polarization case, but the fit requires a very strong magnetic field of approximately 500 G.}}\label{fig9}
\end{figure}

\begin{figure}[htb]
  \begin{center}
    \includegraphics[width=\linewidth]{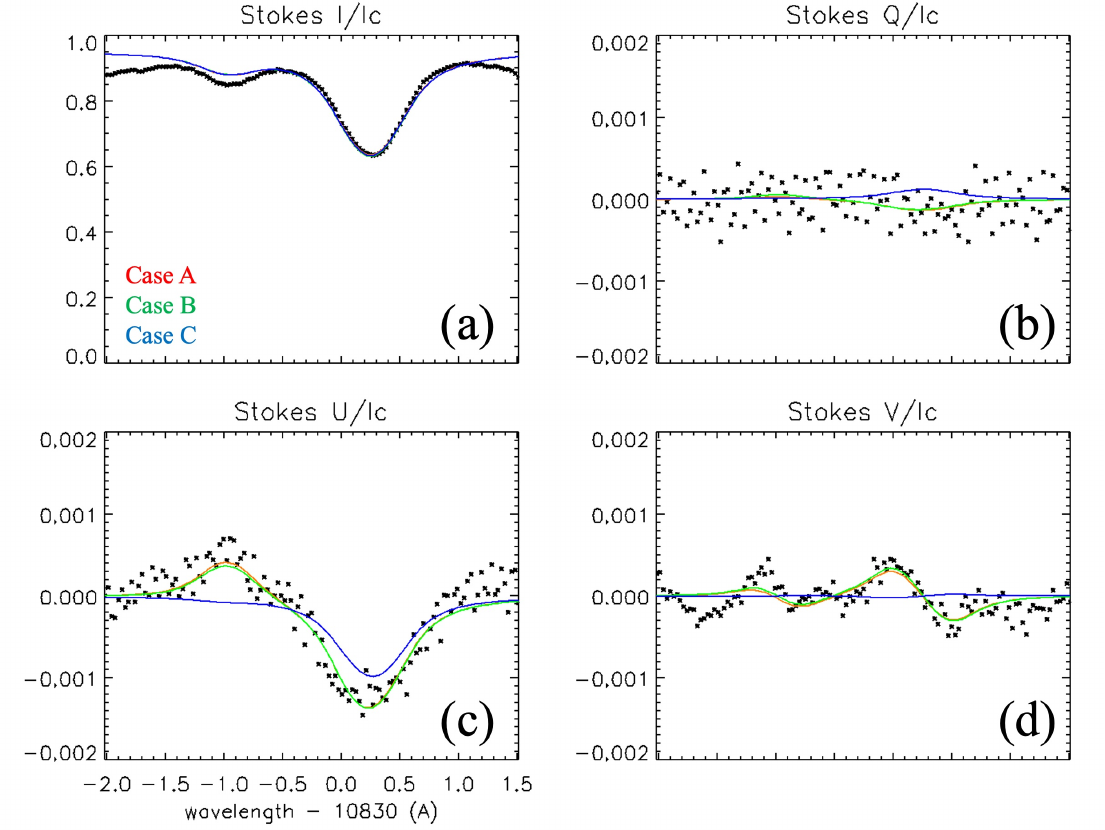}
  \end{center}
  \caption{Stokes profiles of I, Q, U, and V at pixel IV (see the yellow symbol in figure \ref{fig3}a). The black symbols show the observational data. The red, green, and blue solid lines show the full atomic polarization fitting results for ``case A,'' ``case B,'' and ``case C,'' respectively. Note that the continuum level used in normalization of intensity is determined by that of the disk center. \textbf{Alt text: Four panels of observed and modeled Stokes profiles at pixel IV in the filament. Unlike pixel III, the Stokes U profile does not show a double-peaked structure. The observed profiles are generally reproduced by the full atomic polarization inversion for cases A and B.}}\label{fig10}
\end{figure}

As we described in Section \ref{subsec:inv}, we found Zeeman-like profiles in some particular pixels in our observation.
In this section, we present the inversion results for pixels III and IV marked in Figure \ref{fig3} (a).
The Stokes profiles at pixel III exhibit a clear Zeeman-like structure.
Note that the double-peaked profiles in Stokes $U$ were found in the region marked by orange contour shown in panels (b), (c), (d), (e), and (f) in Figure \ref{fig3}.
While the profile at pixel IV is presented for a comparison, $i.e.,$ no Zeeman-like feature is observed in its linear polarization despite its close proximity to pixel III.
In Figure \ref{fig8}, we show the observed Stokes profiles and results of the HAZEL inversion with the full atomic polarization for pixel III marked by the red symbol in Figure \ref{fig3} (a).
Black symbols in the Figure \ref{fig8} show the observed Stokes profiles.
In Figure \ref{fig8} panels (b), (c), and (d), we find $0.1\%$ of the Stokes $Q$ signal, $-0.2\%$ of Stokes $U$ signal, and $\pm0.3\%$ of Stokes $V$ signal in observed profiles.
We found that the shape of the Stokes $U$ profile in pixel III has a double peak in its red component. 
\\
~
The red, green and blue solid curves show the results of the HAZEL inversion for the ``case A'', ``case B'', and ``case C'', respectively.
As shown in panel (a), all the cases successfully fit the observed intensity profile.
First, ``case C'' is rejected because it fails to reproduce the observed Stokes $V$ profiles.
Regarding the fitting of Stokes $U$ and $V$ profiles shown in panels (c) and (d), we find similar fitting profiles in ``case A'' and ``case B'', although neither ``case A'' nor ``case B'' provides a satisfactory fit to the double-peaked Stokes $U$ profile.
The physical parameters at pixel III obtained from the inversion for three cases are summarized in Table \ref{tab3}.
\\
~
Given the possibility that the observed Stokes $U$ profile shown in Figure \ref{fig8} (c) is explained by Zeeman-induced polarization, we performed another Stokes inversion assuming reduced atomic polarization, following the approach of \citet{DiazBaso2016}.
In this case, the contribution of atomic level polarization and the Hanle effect is suppressed, such that the emergent polarization is dominated by the Zeeman effect.
The resulting inversion profiles obtained under this assumption are presented with the orange line in Figure \ref{fig9}.
In panel (c) of Figure \ref{fig9}, we find that the double-peaked Stokes $U$ profile is well fitted compared to the full atomic polarization inversion result shown in panel (c) of Figure \ref{fig8}.
The physical parameters obtained from this inversion are summarized in the last row of the Table \ref{tab3}.
Remarkably, it yields a significantly strong magnetic field at 480 G.
\\
~
In Figure \ref{fig10}, we show the observed Stokes profiles and result of the HAZEL inversion with the full atomic polarization for pixel IV marked by the yellow symbol in Figure \ref{fig3} (a).
Although pixel IV is spatially close to pixel III, we do not find the double-peaked profiles in Stokes $U$ unlike pixel III (see Figure \ref{fig10} (c)).
In panels (b), (c), and (d) of Figure \ref{fig10}, we find $0.05\%$ of the Stokes $Q$ signal, $-0.15\%$ of Stokes $U$ signal, and $\pm0.05\%$ of Stokes $V$ signal in observed profiles.
Regarding the linear polarization signals, the red and blue components of the Stokes $U$ show the opposite sign, suggesting that these signals are caused by the Hanle effect \citep{TrujilloBueno2002}.
In the case of pixel IV, as shown in Figure \ref{fig10}, the full atomic polarization inversion reproduces the observed profiles well except the ``case C''.
We summarized the obtained physical parameters for pixel IV in Table \ref{tab4}.

\begin{table}
\begin{threeparttable}[htb]
  \caption{Inversion results for three cases at pixel III with full atomic polarization cases A, B, C, and reduced atomic polarization case}
  \begin{center}
    \label{tab3}
    \begin{tabular}{lrrrrrrrr}
      \hline
                 & $|\bm{B}|$ & $\theta$ & $\phi$ & $\tau$  & $v_{\mathrm{Dop}}$ & $v_{\mathrm{turb}}$ & $a$    & $ff$ \tnote{1}  \\
                 & [G]     & [$^\circ$]    & [$^\circ$]  &        & [km/s]           & [km/s]            &        &        \\
      \hline
      A           & $44$  & $21$  & $38$  & $1.9$ & $-3$ & $4$ & $1.0$ & $1.0$  \\
      B           & $54$  & $67$  & $35$  & $1.9$ & $-3$ & $4$ & $1.0$ & $1.0$  \\
      C           & $1$   & $132$ & $174$ & $1.9$ & $-3$ & $4$ & $1.0$ & $1.0$  \\
      R \tnote{2} & $480$ & $107$ & $55$  & $1.0$ & $-3$ & $6$ & $0.8$ & $1.0$  \\
      \hline
    \end{tabular}
    \begin{tablenotes}
    \item[1] The filling factor was fixed at unity for all cases.
    \item[2] R denotes the reduced atomic polarization case.  
    \end{tablenotes}
  \end{center}
\end{threeparttable}
\end{table}

\begin{threeparttable}[htb]
  \caption{Inversion results for three cases at pixel IV with full atomic polarization case}
  \begin{center}
    \label{tab4}
    \begin{tabular}{lrrrrrrrr}
      \hline
                 & $|\bm{B}|$ & $\theta$ & $\phi$ & $\tau$  & $v_{\mathrm{Dop}}$ & $v_{\mathrm{turb}}$ & $a$    & $ff$ \tnote{1}  \\
                 & [G]     & [$^\circ$]    & [$^\circ$]  &        & [km/s]           & [km/s]            &        &        \\
      \hline
      A & $9$  & $26$  & $48$  & $0.8$ & $-2$ & $6$ & $1.0$ & $1.0$  \\
      B & $11$ & $70$  & $26$  & $0.8$ & $-2$ & $6$ & $1.0$ & $1.0$  \\
      C & $1$  & $125$ & $180$ & $0.8$ & $-2$ & $6$ & $1.0$ & $1.0$  \\
      \hline
    \end{tabular}
    \begin{tablenotes}
      \item[1] The filling factor was fixed at unity for all cases.
    \end{tablenotes}
  \end{center}
\end{threeparttable}

\clearpage
\section{Discussion}\label{sec:dis}
\subsection{Magnetic field strength and configuration}\label{sec:5.1}
\begin{figure}[htb]
  \begin{center}
    \includegraphics[width=\linewidth]{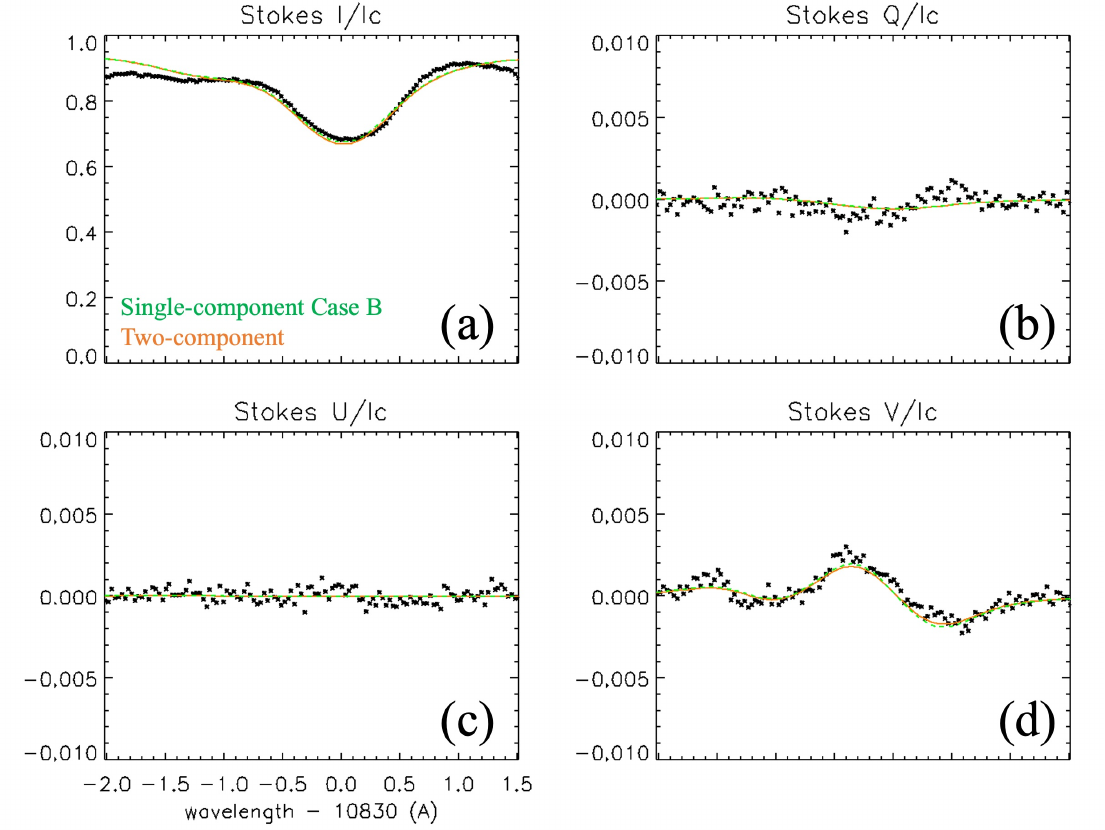}
  \end{center}
  \caption{Observed and fitted Stokes profiles at pixel I comparing the single- and two-component inversions. The green dashed line and orange solid line show the single-component ``case B'' and the two-component fitting results, respectively. The wavelength range is limited around He I 10830 \AA. \textbf{Alt text: Same as Figure \ref{fig4}. Comparison between single-component and two-component inversion fits for pixel I. The two-component model reproduces the observed Stokes profiles using a weak magnetic field in the filament component and a stronger underlying photospheric component.}}\label{fig12}
\end{figure}

\begin{figure}[htb]
  \begin{center}
    \includegraphics[width=\linewidth]{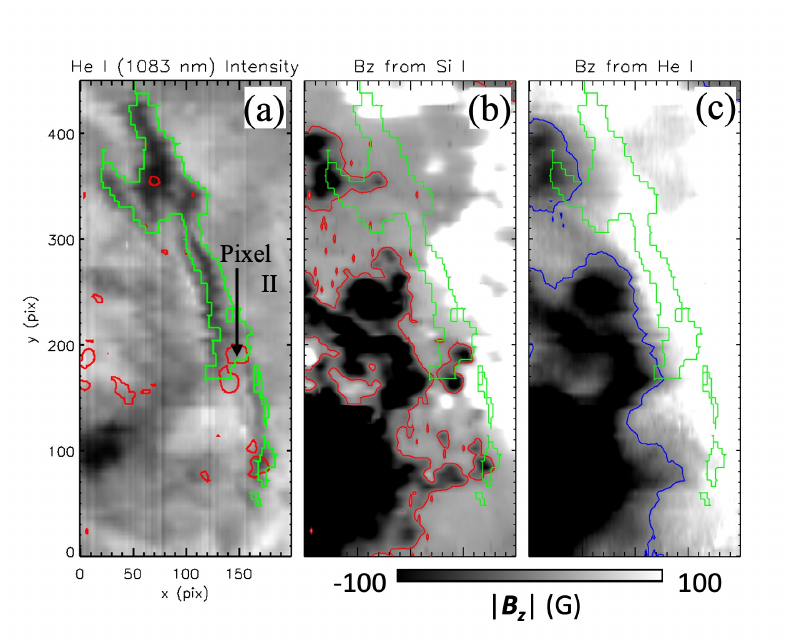}
  \end{center}
  \caption{(a) Intensity map of He I 10830 \AA. Red contour corresponds to the border of region where the signs of line-of-sight component of magnetic field different from Si I and He I. Green contour shows the border of the filament. (b) Line-of-sight component of magnetic field obtained from Si I. Red and green contour show the PIL of Si I and border of the filament, respectively. (c) Line-of-sight component of magnetic field obtained from He I. Blue, and green contour correspond to PIL of He I and border of the filament, respectively. \textbf{Alt text: Three panels are displayed for comparison of line-of-sight magnetic fields derived from He I 10830 \AA~and Si I 10827 \AA~observations. The left panel shows the target filament observed in He I 10830 \AA~line core, and the regions outlined by the red contour indicate areas where the Stokes V signs differ between the chromospheric and photospheric measurements. The middle and the right panels correspond to the Si I and He I magnetic field maps, respectively}}\label{fig11}
\end{figure}

\begin{figure}[htb]
  \begin{center}
    \includegraphics[width=\linewidth]{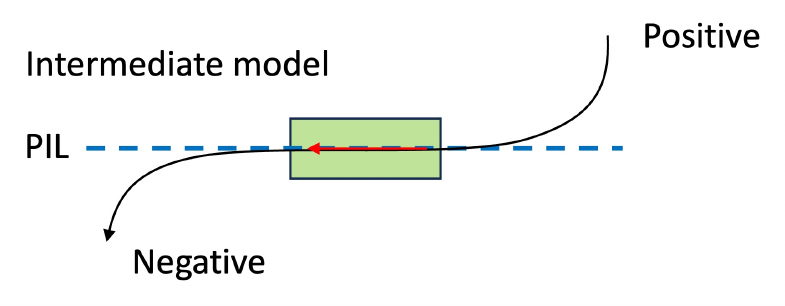}
  \end{center}
  \caption{Schematics of filament magnetic field configuration. Top view of intermediate model. Black arrow, green box, and blue dashed line correspond to magnetic field, plasma material of the filament, and polarity inversion line, respectively. \textbf{Alt text: Schematic diagram illustrating an intermediate filament magnetic configuration in which the horizontal magnetic field is nearly parallel to the filament axis. In this geometry, the distinction between normal-polarity and reverse-polarity models becomes physically ambiguous.}}\label{fig13}
\end{figure}

As described in Section \ref{subsec:invres}, we adopt the results from the full atomic polarization inversion in ``case B'' with the selected magnetic azimuth for pixels I and II, as the most plausible representation of the filament magnetic field.
Such successful inversions were performed for the most part of the dark filament except a small portion in which we found the Zeeman-like profiles.
The exceptional case, $i.e.,$ double-peaked Stokes $U$ profile found in pixel III, will be discussed in the next subsection.
In the following discussion, we focus on the large-scale magnetic properties inferred from the well-fitted regions.
The magnetic field strength obtained from this case is 101 $\pm$ 33 $\mathrm{G}$ (mean $\pm$ 1$\sigma$) within the filament region.
The magnetic field strength of approximately 100 G inferred for the target filament is consistent with previous observational studies of AR filaments ($e.g.$, \cite{Sasso2011,Sasso2014}).
\\
~
\citet{DiazBaso2016} suggested that magnetic field strengths of several hundred gauss can be overestimated when using single-component inversions, owing to the contribution from the underlying magnetic field along the line of sight.
Although the field strength of the filament in this study is much smaller than that, this could also apply to the present situation since Si I 10827 \AA~line infers a field strength of $\sim250$ G in underlying photosphere.
To examine this possibility, we performed a two-component inversion for pixel I, following the approach of \citet{DiazBaso2016}, in which the magnetic field of the underlying component was fixed to the value inferred from the simultaneous Si I 10827 \AA~observations.
We find that the observed Stokes profiles can be reproduced even when the magnetic field strength of the filament component is as low as 10 G.
The resulting fits and the corresponding magnetic field parameters are presented in Figure \ref{fig12} and Table \ref{tab5}, respectively.
Therefore, at this point, we cannot rule out the possibility that the field strengths of active region filaments are even lower than previously reported.
\\
~
In contrast, for pixel II, the line-of-sight magnetic field inferred from the He I and Si I lines shows opposite signs, indicating that the observed signals cannot be explained by a simple superposition of the filament and the underlying photospheric magnetic field (see Figures \ref{fig4} (d) and \ref{fig11} (a)).
The region outlined by the red contour in Figure \ref{fig11} (a) corresponds to areas where the Stokes $V$ signals derived from the Si I and He I lines exhibit opposite signs.
It is evident that several such regions are present within the filament.
Panels (b) and (c) of Figure \ref{fig11} show the spatial distributions of the line-of-sight magnetic field inferred from the Si I and He I lines, respectively.
The Si I map reveals finer spatial structures compared to that obtained from the He I line.
These results suggest that there exist regions where the He I signals are not dominated by the contribution from the underlying background magnetic field.
Moreover, the magnetic field strength of approximately 100 G obtained in this work is in good agreement with measurements of off-limb AR prominences reported by \citet{Hashimoto2026}.
This consistency implies that the inferred field is unlikely to be biased by the contribution of the underlying background magnetic field.
\\
~ Regarding the magnetic field configuration, the direction of the horizontal component is found to be nearly parallel to the filament axis (see panel (f) in Figure \ref{fig7}).
It suggests that there is not much physical significance to distinguish between the normal-polarity and reverse-polarity models as both imply nearly identical magnetic geometries (refer to the schematics shown in Figure \ref{fig13}).

\begin{table*}[htb]
\begin{threeparttable}
  \caption{Inversion results of single- and two-component fitting at pixel I with full atomic polarization case}
  \begin{center}
    \label{tab5}
    \begin{tabular}{lrrrrrrrr}
      \hline
                 & $|\bm{B}|$ & $\theta$ & $\phi$ & $\tau$  & $v_{\mathrm{Dop}}$ & $v_{\mathrm{turb}}$ & $a$    & $ff$ \tnote{1}  \\
                 & [G]     & [$^\circ$]    & [$^\circ$]  &        & [km/s]           & [km/s]            &        &        \\
      \hline
      Single-component (case B) & $109$ & $66$  & $0$  & $0.6$  & $-8$ & $9$ & $1.0$ & $1.0$  \\
      Two-component (upper)     & $10$  & $66$  & $0$  & $0.4$  & $-8$ & $9$ & $1.0$ & $1.0$  \\
      Two-component (lower)     & $250$ & $45$  & $45$ & $0.25$ & $-8$ & $9$ & $1.0$ & $1.0$  \\
      \hline
    \end{tabular}
    \begin{tablenotes}
      \item[1] The filling factor was fixed at unity for all cases.
    \end{tablenotes}
  \end{center}
\end{threeparttable}
\end{table*}

\subsection{Interpretation of Zeeman-like profile}\label{subsec:5.2}
In our spectropolarimetric observation of the AR filament, we found the Zeeman-like profile in Stokes $U$, $i.e.,$ double-peaked symmetric profile, in the red component of the He I 10830 \AA.
As presented in Section \ref{subsec:zeeman}, we performed Stokes inversions on such profiles of He I 10830 \AA~by assuming the full and the reduced atomic polarization with the HAZEL code.
The reduced atomic polarization inversion was performed as a diagnostic experiment, motivated by the expectation that multiple scattering substantially weakens atomic polarization in optically thick regions due to the reduced anisotropy of the radiation field, an effect that is not self-consistently treated in the current version of HAZEL.
In this sense, the reduced atomic polarization inversion may be regarded as an extreme limiting case of the reduction in radiation-field anisotropy found by \citet{VicenteArevalo2023}.
In this inversion, the derived magnetic field was $\sim500$ $\mathrm{G}$, which is comparable to the results reported by \citet{Kuckein2009,Xu2012}.
\citet{DiazBaso2016} interpreted such magnetic field strengths of several hundred gauss as originating primarily from photospheric contributions, rather than from the filament plasma itself.
However, in our case, the Si I 10827 \AA~line observed simultaneously in the same pixel yielded a magnetic field strength of less than 100 $\mathrm{G}$.
Thus, the Zeeman-like profiles cannot be explained by the contribution of the strong magnetic field in the layer beneath the filament as argued by \citet{DiazBaso2016}.
We think that the double-peaked Stokes $U$ profiles observed in our filament are not produced by the Zeeman effect, because core and wings of the He I 10830 \AA~line does not show opposite sign of Stokes $U$ as expected in Zeeman effect and observed by \citet{Kuckein2009}.
\\
~
We further examined the possibility that the observed profiles arise from a superposition of components with different Doppler shifts along the line of sight.
As shown in Figure \ref{fig8}, we found the wavelength separetion between the two peaks in the Zeeman-like Stokes profile of $\sim0.5$ \AA, and it corresponds to $\pm10$ km/s of the line-of-sight velocity, which is comparable with the velocity of counter-streaming flow shown in panel (e) of Figure \ref{fig3}.
However, we think that the double-peak Stokes $U$ profiles cannot be attributed to the counter-streaming flow for a few reasons.
The first reason is that, as shown in Figure \ref{fig8} (c), the blue component, which is expected to be relatively optically thin and therefore more susceptible to superposition effects, does not exhibit a double-peaked structure in the polarization profile.
The second reason is that as shown in Figure \ref{fig8} (d), we do not find the feature which shows the superposion of different line-of-sight velocity components in Stokes $V$ profile, $i.e.,$ the Stokes $V$ profile shown in panel (d) of Figure \ref{fig8} shows single pair of positive and negative polarization signal as typically observed in the presence of the Zeeman effect with a single line-of-sight velocity.
The third reason is that as shown in Figure \ref{fig3} (e), the region indicated by the orange contour including pixel III, where the Zeeman-like Stokes profiles are detected, is located within a region dominated by blueshifted velocities, apart from the boundary of blue and red shifted regions.
Therefore, we suggest that the observed Zeeman-like profiles cannot be accounted for by a superposition of redshifted and blueshifted components, reinforcing the conclusion that an alternative physical mechanism is required to explain the Zeeman-like profiles.
\\
~
Remarkably, the Zeeman-like Stokes profiles were found in optically thick regions at the core of He~I 10830~\AA~(see Figure \ref{fig3} (f)).
The optical thickness inferred from the HAZEL inversion ranges from approximately 1.4 to 2.5, with a mean value of about 2.0.
Although the optical thickness inferred by HAZEL should not be regarded as quantitatively robust in pixels where the observed Zeeman-like Stokes profiles are not satisfactorily reproduced, neither should it be regarded as entirely unreliable.
The optical thickness is primarily constrained by the Stokes~I profile, while the linear polarization signals have amplitudes of only the order of $10^{-3}$ relative to the continuum intensity.
Therefore, although the imperfect reproduction of the polarization profiles indicates limitations of the current inversion framework, the good agreement obtained for Stokes~I suggests that the inferred optical thickness still provides a reasonable estimate of the opacity of the observed filament.
\\
~
\citet{VicenteArevalo2023} demonstrated that, once the slab and flat-spectrum approximations adopted in HAZEL break down in optically thick regions, the individual components of the He~I 10830~\AA\ multiplet experience different radiation fields owing to self-consistent radiative transfer.
Consequently, the anisotropy responsible for atomic polarization is no longer described by the simple assumptions adopted in HAZEL.
However, their study was based on the CRD, and they did not specifically reproduce the wavelength-dependent depolarization pattern, depression at the line center.
\\
~
It is recalled that as discussed by \citet{Heinzel1981} and \citet{Hubeny2015}, the frequency redistribution of scattered photons becomes important when the spectral line width is primarily determined by thermal Doppler broadening and the scattering process is considered in the observer's frame.
Under such conditions, partial frequency redistribution (PRD) is expected to introduce different polarization properties between the line core and the wings.
Therefore, although the present observations do not demonstrate that PRD is uniquely responsible for the observed Zeeman-like profiles, they suggest that radiative-transfer calculations incorporating both the breakdown of the HAZEL approximations in optically thick regions and PRD may provide a more realistic description of the observed wavelength-dependent depolarization.

\section{Summary}
We performed spectropolarimetric observations of an active-region filament in NOAA 13092 using the He~I 10830~\AA\ triplet and the Si~I 10827~\AA\ line obtained with the Domeless Solar Telescope at Hida Observatory. The full atomic polarization inversion with the HAZEL code indicates that the filament magnetic field has a typical strength of approximately 100 G and is predominantly horizontal, with its direction nearly parallel to the filament axis. Consequently, the magnetic field orientation does not allow us to distinguish between the classical normal- and reverse-polarity configurations.
\\
~
We also identified localized Zeeman-like Stokes~$U$ profiles that could not be satisfactorily reproduced by the current HAZEL inversion framework. These profiles were preferentially found in relatively optically thick regions of the filament. Our discussion suggests that the discrepancy primarily reflects the breakdown of the simplifying assumptions adopted in HAZEL in optically thick media, as pointed out by \citet{VicenteArevalo2023}. Furthermore, the observed wavelength dependence of the depolarization, namely stronger depolarization at the line center than in the wings, may require more sophisticated radiative-transfer modeling that includes not only self-consistent radiation fields and differential illumination of the multiplet components, but also possibly partial frequency redistribution. The present observations therefore provide new observational constraints for future modeling of scattering polarization in optically thick active-region filaments.



\section*{Funding}
  This work was supported by JSPS KAKENHI Grant Number JP23K19078 and JP26K17213.
  This research was also supported by a grant from the Hayakawa Satio Fund awarded by the Astronomical Society of Japan.

\begin{ack}
  The authors thank all of the staff members at Hida observatory for continuous support for our observations.
  We also thank Dr. KD Leka for significant help for resolving 180 deg ambiguity.  
  SDO is a mission of NASA's Living With a Star Program.
\end{ack}









\end{document}